\documentstyle[12pt,epsf]{article}
\begin{document}
\begin{titlepage}
  \begin{flushright}
    KUNS-1671\\[-1mm]
    YITP-00-43\\[-1mm]
    hep-ph/0008120
  \end{flushright}
  \begin{center}
    \vspace*{1.4cm}
    
  {\Large\bf Fermion Mass Hierarchies and Small Mixing Angles\\
    from Extra Dimensions}\vspace*{1cm}
  
  Masako Bando\footnote{E-mail: bando@aichi-u.ac.jp},
    Tatsuo Kobayashi\footnote{E-mail: kobayash@gauge.scphys.kyoto-u.ac.jp},
    Tatsuya Noguchi\footnote{E-mail: noguchi@gauge.scphys.kyoto-u.ac.jp} and
    Koichi Yoshioka\footnote{E-mail: yoshioka@yukawa.kyoto-u.ac.jp}
  \vspace*{5mm}
  
  $^*${\it Aichi University, Aichi 470-0296, Japan}\\
  $^{\dagger,\ddagger}${\it Department of Physics, Kyoto
    University, Kyoto 606-8502, Japan}\\
  $^{\protect\S}${\it Yukawa Institute for Theoretical Physics, Kyoto
  University, Kyoto 606-8502, Japan} 
  \vspace{1.2cm}
  
  \begin{abstract} 
    In this paper we study renormalization-group evolutions of Yukawa
    matrices enhanced by Kaluza-Klein excited modes and analyze their
    infrared fixed-point structure. We derive necessary conditions to
    obtain hierarchies between generations on the fixed point. These
    conditions restrict how the fields in the models can extend to
    higher dimension. Several specific mechanisms to realize the
    conditions are presented. We also take account of generation
    mixing effects and find a scenario where the mixing angles become
    small at low energy even with large initial values at high-energy
    scale.  A toy model is shown to lead realistic quark mass matrices.
  \end{abstract}
\end{center}
\end{titlepage}

\setcounter{footnote}{0}

\section{Introduction}
\setcounter{equation}{0}

The fermion mass hierarchy problem has been one of the most
challenging problems in particle physics: the observed quarks and
leptons show hierarchical mass difference between three
generations. In the Standard Model, their particle masses are only
input parameters.  However, from the viewpoint of the unification
scenario, this fermion mass hierarchy should also be explained. A
number of ideas have been proposed so far. Among them the
Froggatt-Nielsen mechanism~\cite{FN} is one of possibilities which can
explain the power hierarchy of Yukawa couplings via higher-dimensional
operators constrained by some extra symmetries. Unfortunately,
however, there is no principle to determine the coefficients of these
operators and we can at most predict the order of Yukawa
couplings. The coefficients could be determined by some more
fundamental principle.

Recently, models with extra spatial dimensions have been
investigated~\cite{antoniadis,ED}, and they have provided novel approaches to
gauge coupling unification~\cite{gut}, fermion masses~\cite{mass},
possibilities of the experimental observations~\cite{exp} and so
forth. In these models, the existence of Kaluza-Klein (KK) excited
modes plays an important role: because of the contribution from the
towers of KK modes (of the Standard Model fields), the
renormalization-group equations (RGE) of gauge and Yukawa couplings
have ``power-law'' dependence on energy scale~\cite{power,gut}. The
emergence of the power-law running behavior is very attractive when it
comes to discussing the fermion mass hierarchy.

In our previous paper~\cite{BKNY}, we have shown that Yukawa couplings
from higher-dimensional models can be highly suppressed under RG
evolution with the value of coupling being stabilized as an infrared
stable fixed point. Furthermore, extending the analysis to the case
with two Yukawa couplings, we have shown that hierarchically different
values can be realized as infrared fixed point values.

The aim of this paper is to study further the power-law running effect
on Yukawa couplings and its infrared fixed-point structure in more
detail. We give the necessary conditions for field configurations to
yield mass hierarchies, and propose some mechanism realizing the
conditions. That gives the hint how to arrange the matter, Higgs and
gauge fields in the extra dimensions.  Although in this paper, we do
not perform explicit model-buildings, the obtained conditions are
useful in constructing realistic higher-dimensional models.

The article is organized as follows. In section 2, we discuss the
power-law running behavior of gauge and Yukawa couplings and their
infrared fixed-point structure in the case with no generation
mixing. In section 3, we study under what conditions the Yukawa
beta-functions are
made generation-dependent and hierarchical Yukawa
couplings are realized. We investigate RG flows of Yukawa couplings in
section 4, taking account of generation-mixing and give a toy model which
reproduces realistic quark matrices. Section 5 is devoted to the
summary and discussion.

\section{RGE running of Yukawa couplings}
\setcounter{equation}{0}

In this section, we study power-law running behavior of gauge and
Yukawa couplings and their fixed-point structure in supersymmetric
(SUSY) models with extra spatial dimensions. 
We assume that matter, Higgs and gauge fields
can extend into the extra dimensions other than our four-dimensional
spacetime. Models with extra dimensions generally involve different
compactification scales depending on their directions, but 
for simplicity, we here take all of these scales as the universal
compactification scale $\mu_0$, above which KK modes appear. Below
$\mu_0$, the KK modes decouple and the models become four-dimensional
$N=1$ SUSY theories with only light modes. Thus, the infrared fixed
point values at $\mu_0$ are regarded
as the boundary values at $\mu_0$ of ordinary $N=1$ SUSY theory. 
Our aim is to realize hierarchical values of Yukawa couplings at the scale $\mu_0$.

Let us consider the case with one gauge coupling constant $g$, which
may be identified with the $SU(3)$ gauge coupling in the SUSY standard
models. We consider an $N=1$ SUSY gauge-Yukawa system with Yukawa
couplings $y_i$,
\begin{eqnarray}
  W &=& \sum_i y_i {\Psi_L}_i {\Psi_R}_i H,
\end{eqnarray}
where ${\Psi_L}_i$ and ${\Psi_R}_i$ are the $i$-th generation of
matter superfields. Here we have neglected generation mixing.

\begin{figure}[htbp]
\hspace*{.18\linewidth}
\parbox{.4\linewidth}{
  \epsfbox{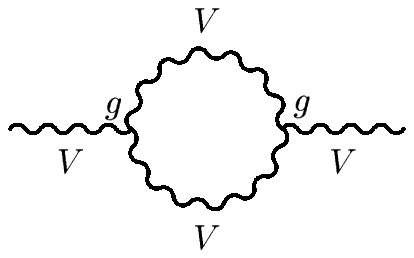}\hspace*{0.255\linewidth}(a)}
\parbox{.4\linewidth}{
  \epsfbox{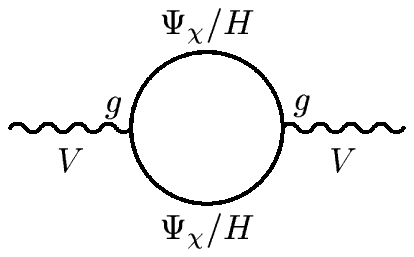}\hspace*{0.255\linewidth}(b)}
    \caption{The Feynman diagrams contributing to $\gamma_g$. $\chi$
    represents $L$ and $R$ in the diagrams (b).}
  \label{fig:gauge}
\vspace{0.1\linewidth}

\hspace*{.18\linewidth}
\parbox{.4\linewidth}{
  \epsfbox{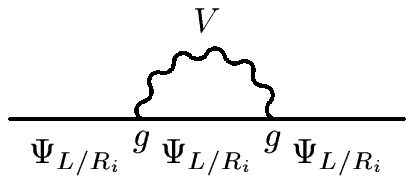}\hspace*{0.255\linewidth}(a)}
\parbox{.4\linewidth}{
  \epsfbox{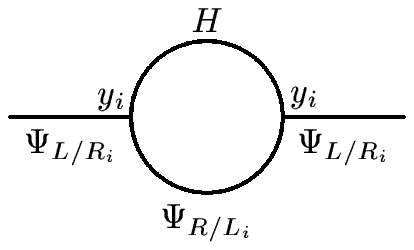}\hspace*{0.255\linewidth}(b)}
    \caption{The Feynman diagrams contributing to $\gamma_{L/Ri}$.}
  \label{fig:matter}
\vspace{0.1\linewidth}

\hspace*{.18\linewidth}
\parbox{.4\linewidth}{
  \epsfbox{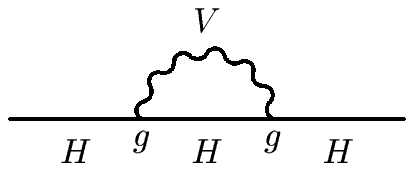}\hspace*{0.255\linewidth}(a)}
\parbox{.4\linewidth}{
  \epsfbox{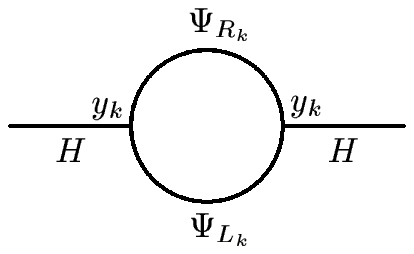}\hspace*{0.255\linewidth}(b)}
    \caption{The Feynman diagrams contributing to $\gamma_H$. The
    index $k$ runs over all generations in the diagrams (b).}
  \label{fig:higgs}
\end{figure}

In this setup, the one-loop beta-functions of gauge and Yukawa couplings above
$\mu_0$ generally contain power terms of energy scale~\cite{power,gut}
while the beta-functions below $\mu_0$ are controlled by usual
logarithmic terms.\footnote{Higher loop corrections have 
been studied in Ref.~\cite{two-loop}.} 
Above $\mu_0$, the RGE of the gauge coupling
$\alpha\equiv g^2/4\pi$ is written as
\begin{eqnarray}
  \frac{\partial \alpha}{\partial t} &=& -\frac{b}{2\pi}
  \left(\frac{\Lambda}{\mu}\right)^{\delta_g}\alpha^2 +\cdots,
  \qquad\quad t = \ln\biggl(\frac{\Lambda}{\mu}\biggr),
  \label{EDgauge}
\end{eqnarray}
where $\Lambda$ is the cutoff energy scale ($\mu_0 < \mu <\Lambda$)
and $b$ is determined by group-theoretical and phase-space integration
factors. The ellipsis denotes sub-leading terms. The power of $\Lambda
/\mu$, $\delta_g$, is generated by the KK-mode contribution of gauge
fields (Fig.~\ref{fig:gauge}a), matter and Higgs fields
(Fig.~\ref{fig:gauge}b), which propagate in the loops. Note that all
of the external lines in Figs.~\ref{fig:gauge},~\ref{fig:matter}
and~\ref{fig:higgs} are KK zero-modes. The case with no extra
dimension ($\delta_g=0$) corresponds to usual logarithmic behavior in
four dimensions. Similarly the RGEs of the Yukawa couplings
$\alpha_{y_i}\equiv y^2_i/4\pi$ are expressed as
\begin{eqnarray}
  \frac{\partial{\alpha_y}_i}{\partial t} &=&
  \frac{{\alpha_y}_i}{2\pi}\,(\gamma_{L_i}+\gamma_{R_i}+\gamma_{H}),
  \label{EDyukawa}
\end{eqnarray}
where $\gamma_{\chi_i}$ ($\chi = L$ or $R$) and $\gamma_H$ are the
anomalous dimensions of ${\Psi_\chi}_i$ and $H$, respectively:
\begin{eqnarray}
  \gamma_{\chi_i} &=& -a_{\chi_i}
  \left(\frac{\Lambda}{\mu}\right)^{\delta^{\chi_i}_y} {\alpha_y}_i 
  +c_{\chi_i}\left(\frac{\Lambda}{\mu}\right)^{\delta^{\chi_i}_g}
  \alpha +\cdots,
  \label{gammachi} \\
  \gamma_H &=& -\sum_i a_{H_i} 
  \left(\frac{\Lambda}{\mu}\right)^{\delta^H_{y_i}} {\alpha_y}_i
  +c_H\left(\frac{\Lambda}{\mu}\right)^{\delta^H_g} +\cdots
  \label{gammaH}
\end{eqnarray}
with $a$ and $c$ being written in terms of group-theoretical and phase
 factors. In the following, we drop the beta-function coefficients,
 $a$, $b$ and $c$ since only exponents of $\Lambda/\mu$ are important
 for our discussion.\footnote{One can easily restore the omitted
 coefficients, associating them with the exponents $\delta$'s.} The
 exponents $\delta^{\chi_i}$'s in the $i$-th generation anomalous
 dimension $\gamma_{\chi_i}$ reflect the properties of the
 intermediating gauge fields (Fig.~\ref{fig:matter}a) and Higgs fields
 (Fig.~\ref{fig:matter}b). As for the Higgs anomalous dimension
 $\gamma_H$, the exponents are determined by the contributions from
 the gauge fields (Fig.~\ref{fig:higgs}a) and matter fields
 (Fig.~\ref{fig:higgs}b). Note that $\gamma_H$ includes the
 contributions from all generations.

Let us consider the two-generation case for concreteness. The RGEs of
the Yukawa couplings ${\alpha_y}_1$ and ${\alpha_y}_2$ are given by
\begin{eqnarray}
  \frac{d}{dt}\left(\frac{\alpha_{y_1}}{\alpha}\right) &=&
  \frac{1}{2\pi}\left[-\left(\frac{\Lambda}{\mu}\right)^{\delta^y_1}
    \frac{\alpha_{y_1}}{\alpha}
    -\left(\frac{\Lambda}{\mu}\right)^{\delta^H_{y_2}} 
    \frac{\alpha_{y_2}}{\alpha}
    +\left(\frac{\Lambda}{\mu}\right)^{\delta^g_1}\right]\alpha_{y_1},
  \label{RG-a1} \\
  \frac{d}{dt}\left(\frac{\alpha_{y_2}}{\alpha}\right) &=&
  \frac{1}{2\pi}\left[-\left(\frac{\Lambda}{\mu}\right)^{\delta^H_{y_1}}
    \frac{\alpha_{y_1}}{\alpha}
    -\left(\frac{\Lambda}{\mu}\right)^{\delta^y_2}
    \frac{\alpha_{y_2}}{\alpha} 
    +\left(\frac{\Lambda}{\mu}\right)^{\delta^g_2}\right]\alpha_{y_2},
  \label{RG-a2}
\end{eqnarray}
where $\delta^y_i\equiv {\rm Max}\,(\delta^{\chi_i}_y,\delta^H_{y_i})$
and $\delta^g_i\equiv {\rm Max}\,(\delta_g,\delta^{\chi_i}_g,\delta^H_g)$. 
The fixed-point solutions at $\mu=\mu_0$ for $\alpha_{y_1}/\alpha$ and 
$\alpha_{y_2}/\alpha$ satisfy the 
equations $\frac{d (\alpha_{y_i}/\alpha)}{dt}\Bigr|_{\mu=\mu_0}=0$, i.\/e.\/,
\begin{eqnarray}
  \pmatrix{\left(\Lambda/\mu_0\right)^{\delta_1^y} & 
    \left(\Lambda/\mu_0\right)^{\delta_{y_2}^H} \cr
    \left(\Lambda/\mu_0\right)^{\delta_{y_1}^{H}} & 
    \left(\Lambda/\mu_0\right)^{\delta_{2}^y}}
  \pmatrix{ ({\alpha_y}_1/\alpha)^* \cr ({\alpha_y}_2/\alpha)^*} \;=\; 
  \pmatrix{\left(\Lambda/\mu_0\right)^{\delta_1^g} \cr 
    \left(\Lambda/\mu_0\right)^{\delta_2^g}}.
  \label{fpeq}
\end{eqnarray}
Then the solutions are found as,
\begin{eqnarray}
  \left(\frac{{\alpha_y}_1}{\alpha}\right)^* &=& D^{-1} \left[
    \left(\frac{\Lambda}{\mu_0}\right)^{\delta^g_1+\delta^y_2}
    -\left(\frac{\Lambda}{\mu_0}\right)^{\delta^g_2+\delta^H_{y_2}}
  \right], 
  \label{alpha1}\\
   \left(\frac{{\alpha_y}_2}{\alpha}\right)^* &=& D^{-1} \left[
    \left(\frac{\Lambda}{\mu_0}\right)^{\delta^g_2+\delta^y_1}
    -\left(\frac{\Lambda}{\mu_0}\right)^{\delta^g_1+\delta^H_{y_1}}
 \right],
  \label{alpha2}
\end{eqnarray}
with the determinant of the matrix on the left hand side of
Eq.~(\ref{fpeq}): 
\begin{eqnarray}
  D &=& \left(\frac{\Lambda}{\mu_0}\right)^{\delta^y_1+\delta^y_2} 
  -\left(\frac{\Lambda}{\mu_0}\right)^{\delta^H_{y_1}+\delta^H_{y_2}}.
  \label{det}
\end{eqnarray}
Since the determinant $D$, which is non-negative by definition, must be
``positive'' in order to have definite solutions, the following
condition must be satisfied\\ 
\noindent
(Condition I)\\[-7.5ex]
\begin{equation}
\delta^y_1+\delta^y_2 \;>\; \delta^H_{y_1}+\delta^H_{y_2}.\label{det-cond}
\end{equation}

Condition I implies that at least one of
$\delta^{{\chi_i}}$ must be larger than $\delta^H_{y_i}$.
Moreover, the requirement that the solutions~(\ref{alpha1}) and (\ref{alpha2}) should be positive imposes additional constraints
for $\delta$'s,\\
\\
\noindent
(Condition II)\\[-11.0ex]
\begin{eqnarray}
  \delta^g_1+\delta^y_2 &>& \delta^g_2+ \delta^H_{y_2}, 
  \label{pos-cond1}\\
  \delta^g_2+\delta^y_1 &>& \delta^g_1+ \delta^H_{y_1}.
  \label{pos-cond2}
\end{eqnarray}
To see the fixed-point solutions more explicitly, let us define $R_i$:
\begin{eqnarray}
  R_1 &=& D \left[
    \left(\frac{\Lambda}{\mu}\right)^{\delta^g_1+\delta^y_2} 
    -\left(\frac{\Lambda}{\mu}\right)^{\delta^g_2+\delta^H_{y_2}}
  \right]^{-1} {{\alpha_y}_1\over\alpha}, \\
  R_2 &=& D \left[
    \left(\frac{\Lambda}{\mu}\right)^{\delta^g_2+\delta^y_1}
    -\left(\frac{\Lambda}{\mu}\right)^{\delta^g_1+\delta^H_{y_1}} 
  \right]^{-1} {{\alpha_y}_2\over\alpha}.
\end{eqnarray}
Under Condition I, we have $D\simeq (\Lambda/\mu)^{\delta^y_1 +
  \delta^y_2}$ and their RGEs are written 
\begin{eqnarray}
  \frac{dR_1}{dt} &=& -\frac{\alpha}{2\pi} 
  \left(\frac{\Lambda}{\mu}\right)^{\delta^g_1} R_1 \left[R_1
    +R_2\left(\frac{\Lambda}{\mu}\right)^{\delta^g_2+\delta^H_{y_2}
      -\delta^y_2-\delta^g_1} -1\right], \label{rge1}\\
  \frac{dR_2}{dt} &=& -\frac{\alpha}{2\pi}
  \left(\frac{\Lambda}{\mu}\right)^{\delta^g_2} R_2 \left[R_1
    \left(\frac{\Lambda}{\mu}\right)^{\delta^g_1+\delta^H_{y_1}
      -\delta^y_1-\delta^g_2} +R_2 -1 \right]. \label{rge2}
\end{eqnarray}
Condition II implies that the $R_2$ ($R_1$) term in the first (second)
equation is suppressed enough than the other terms. We then find from
Eqs.~(\ref{rge1}) and (\ref{rge2}) that $R_i$ converge into
$R_i=1$. The convergence behavior onto their fixed points is important
to regard the fixed-point values as boundary conditions for the
realistic quark Yukawa couplings. If the above two conditions are
satisfied, the RGEs of $R_i$ are solved as 
\begin{eqnarray}
  \frac{R_i(t)-1}{R_i(t)} &=& \xi_i\,\frac{R_i(0)-1}{R_i(0)}, \label{R}
\end{eqnarray}
with $R_i(0)$ being initial values at the cutoff scale $\Lambda$. The
quantity $\xi_i$ is obtained
\begin{eqnarray}
  \xi_i &=&\frac{1}{E_i(t)}\biggl(\frac{\alpha(t)}{\alpha(0)}\biggr),
  \label{xi}
\end{eqnarray}
where
\begin{eqnarray}
  E_i(t) &=& \exp\biggl[\int^t_0 dt'\,\frac{1}{2\pi}
    \left(\frac{\Lambda}{\mu'}\right)^{\delta^g_i}\alpha\biggr].
\end{eqnarray}
The quantity $\xi_i$ measures the rapidity of convergence onto the
fixed points $R_i=1$. In extra-dimensional models, the Yukawa
couplings converge rapidly even in asymptotically-free gauge theories,
if the relation $\delta^g_i>\delta_g$ is satisfied. That is in
contrast to the slow convergence in four-dimensional
asymptotically-free gauge theories. This property is interesting for
constructing asymptotically-free models with strong infrared
convergence~\cite{BKNY}. Note that even in the case of
$\delta^g_i<\delta_g$, strong convergence of Yukawa couplings is
utilized in infrared-free models ($b<0$)~\cite{anf}.

The dashed lines in Figure 4 show typical running behavior of
${\alpha_y}_1$ and ${\alpha_y}_2$ calculated from RGEs (\ref{RG-a1})
and (\ref{RG-a2}). In this figure, we have used the RGEs (\ref{RG-a1})
and (\ref{RG-a2}). The solid lines which correspond to the fixed-point
solutions (\ref{alpha1}) and (\ref{alpha2}).  From the figure, we see
that the good convergence to the fixed points and a large hierarchy
between two couplings are indeed viable.
\begin{figure}[htbp]
  \begin{center}
    \leavevmode
    \epsfxsize=9cm
    \epsfbox{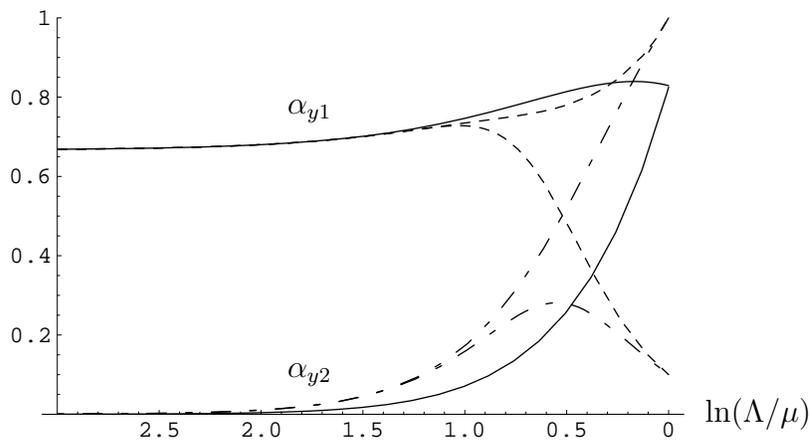}
    \put(8,18){$\ln(\Lambda/\mu)$}
    \put(-150,35){$\alpha_{y2}$}
    \put(-150,135){$\alpha_{y1}$}
    \caption{Typical running behaviors of Yukawa couplings. 
      The exponents in the RGEs are assumed as follows:
      $\delta_g = 2$, $\delta^y_1 = 5$, $\delta^y_2 = 4$, $\delta^h_1 = 3$,
      $\delta^h_2 = 2$, $\delta^g_1 = 2$ and  $\delta^g_2 = 4$.
      The solid lines are fixed-point solutions (that are
      energy-dependent). The dashed and dot-dashed lines denote the
      RGE running of ${\alpha_y}_1$ and ${\alpha_y}_2$, respectively.
      In the figure,
      we take two initial values, ${\alpha_y}_1(0)={\alpha_y}_2(0)=1$
      and $0.1$. }
  \end{center}
\end{figure}

It is straightforward to extend the above discussions to the cases
with more than two generations. First, to obtain definite fixed-point
values, at least one of the Yukawa terms in matter anomalous
dimensions should be larger than that in the Higgs anomalous
dimension, i.\/e.\/\ $\delta^y_i>\delta^H_{y_i}$ (the non-vanishing
determinant condition). Furthermore, the positiveness of fixed-point
values must be satisfied. That requires the condition similar to
(\ref{pos-cond1}), etc. If these conditions are satisfied, Yukawa
couplings ${\alpha_y}_i$ are fixed by the infrared fixed-point values,
${\alpha_y}_i\sim (\mu_0/\Lambda)^{\delta^y_i-\delta^g_i}\alpha$ at
the KK threshold scale $\mu_0$.

\section{Realization for hierarchies}
\setcounter{equation}{0}

We have derived Conditions I and II for realizing hierarchically
suppressed couplings on infrared fixed points of RG evolution. In this
section, we propose several mechanisms to realize the above
conditions.

First, we consider the condition to have generation-{\it dependent}\,
fixed-point values of Yukawa couplings. The beta-function of Yukawa
coupling is given by $\beta\propto y(\gamma_L+\gamma_R+\gamma_H)$. If
the Higgs contribution, $\gamma_H$, is dominant, generation-dependent
terms become sub-dominant since the $\gamma_H$ contribution commonly
appears in all Yukawa beta-functions. A key ingredient for Yukawa
hierarchy is therefore to suppress the Higgs anomalous dimension: the
matter contributions, $\gamma_L$ and/or $\gamma_R$, should be larger
than $\gamma_H$, i.\/e.\/\ $\delta^y>\delta^H_y$. Note that this
condition corresponds to Condition I. There are two possibilities for
suppressing the Higgs effects: (i) the couplings between Higgs
zero-modes and KK modes of $\Psi_L$ and $\Psi_R$ are forbidden and
(ii) the loop diagrams contributing to $\gamma_H$ cancel out. We
describe the models satisfying the condition (i) in sections 3.1 and
3.2, and the condition (ii) in section 3.3.

\subsection{Charge conservation rule}

Generally, each bulk field is allowed to have various values of KK
momentum under the compactification. At interaction vertices, however,
the momentum conservation or some charge-conservation rules associated
with extra dimensions restricts possible combinations of KK modes.  To
put it more precisely, the loop diagrams where two KK modes extend in
common directions of extra dimensions can contribute to wave-function
renormalization.

Now let us consider the Yukawa term $\Psi_L\Psi_R H$. The KK modes
contributing to the anomalous dimension $\gamma_{L\,(R)}$
(Fig.~\ref{fig:matter}b) are those along the directions to which
$\Psi_{R\,(L)}$ and $H$ extend in common. The KK modes with nonzero
charges for other directions cannot propagate. Likewise, in
Fig.~\ref{fig:higgs}b, only the KK modes of $\Psi_L$ and $\Psi_R$
extending into same extra dimensions contribute to the Higgs anomalous
dimension. It is hence required for suppressing $\gamma_H$ to reduce
the number of dimensions that $\Psi_L$ and $\Psi_R$ expand in common.
For example, in the case that there is no common extra dimensions
between $\Psi_L$ and $\Psi_R$, we obtain $\delta^H_y=0$ and the
anomalous dimension $\gamma_H$ reduces to logarithmic behavior. Then,
assuming the Higgs field extends to some numbers of extra dimensions,
we can obtain a large $\delta^L_y$ and/or $\delta^R_y$ whose values
are generation-dependent and thereby induce hierarchies.

To see how this mechanism works, consider a two-generation model with
the configuration shown in Table 1.
\begin{table}[htbp]
  \begin{center}
    \begin{tabular}{c|cccccc}
      & ~5\quad &\quad 6\quad &\quad 7\quad &\quad 8\quad &\quad
      9\quad &\quad 10\quad \\ \hline 
      $H$ & ~$\bigcirc$ &\quad $\bigcirc$ &\quad $\bigcirc$ &\quad
      $\bigcirc$ &\quad $\bigcirc$ &\quad $\bigcirc$ \\
      ${\Psi_L}_1$ & ~$\bigcirc$ & & & & & \\
      ${\Psi_L}_2$ & ~$\bigcirc$ &\quad $\bigcirc$ & & & & \\ 
      ${\Psi_R}_{1,2}$ & & & & & & \\
    \end{tabular}
    \caption{An example of field configuration generating Yukawa
      hierarchy. In the first row, the numbers $5,6,\cdots, 10$
      represent the labels of the extra spatial dimensions. Each
      circle denotes that the corresponding field can propagate
      through there.}
\end{center}
\end{table}
In the table, each circle denotes that the corresponding field
propagates through that direction. With this setup, we have
$\delta^H=0$ because $\Psi_L$ and $\Psi_R$ have no overlapping
dimensions (except for our four dimensions). We also find
$\delta^y_1=1$ and $\delta^y_2=2$. It is interesting to note that
these non-zero powers come from the anomalous dimensions of the
right-handed field, $\gamma_{Ri}$, while those of the left-handed
fields, which spread into extra dimensions, do not contribute to
$\delta^y_i$. The above model thus realizes a hierarchy under the RG
evolutions down to the infrared.

In this scenario, the number of extra dimensions is responsible for
hierarchically different mass parameters. For example, since the
perturbative string theories demand six extra spatial dimensions, it
can generate at most six mass gaps. Another interesting point is that
the `left-right asymmetry' is essential for generating
family-dependences in the beta functions, $\beta_i\propto
y_i(\gamma_{Li}+\gamma_{Ri}+\gamma_H)$. We will apply this left-right
asymmetric scenario to find a realistic form of quark mass matrices in
later section, including generation-mixing effects.

\subsection{$N=2$ hypermultiplet}

Another mechanism satisfying the condition (i) is to utilize $N=2$ (or
larger) supersymmetry. In generic $N=2$ SUSY models, Yukawa couplings
among $N=2$ hypermultiplets are forbidden by supersymmetry.

We suppose that the models consist of two parts. One is a
four-dimensional theory derived from higher dimension by orbifold
($Z_2$) projections. It is noted that the $Z_2$ symmetry is only
responsible for having chiral nature of matter fields and not for
suppressing anomalous dimensions. The other sector is a genuine
four-dimensional theory (defined on the $Z_2$ fixed point). In
addition, we assume that the former part respects $N=2$ SUSY before
the projection and forbids Yukawa interactions among hypermultiplets
in the model. On the other hand, there could exist Yukawa couplings
between the hypermultiplets and $N=1$ superfields on the
four-dimensional theory even with this assumption.

For simplicity, let us again consider a two-generation model with the
Higgs $H$ and the second-generation fields ${\Psi_{L,R}}_2$ coming
from $N=2$ multiplets while ${\Psi_{L,R}}_1$ belong to
four-dimensional $N=1$ multiplets.\footnote{If one assumes the
KK-momentum conservation, either of ${\Psi_{L,R}}_1$ should be
assigned to an $N=2$ hypermultiplet. Otherwise, we have no enhanced
RG-running effect $(\delta^y_1=0)$. It may be, however, natural not to
assume such a conservation rule as long as one exploits orbifold
symmetries.}  In this case, all the Yukawa couplings of the KK modes
of $H$, ${\Psi_L}_2$ and ${\Psi_R}_2$ are not allowed and consequently
we have $\delta^H=\delta^y_2=0$. Since the Yukawa coupling of the
first-generation fields is allowed, we obtain $\delta^y_1\neq 0$ from
the Higgs KK-modes contribution. We thus obtain a Yukawa hierarchy
between the generations. It turns out that the Yukawa coupling
originated from $N=1$ multiplets has power-law RG running while that
from $N=2$ theory runs logarithmically (see Table 2). The order of
hierarchy is determined by the number of extra dimensions in which the
Higgs multiplet lives. Note that with this mechanism, we can
accomplish only one difference between the couplings: the power-law
and logarithmic running behaviors.
\begin{table}[htbp]
  \begin{center}
    \begin{tabular}{c|lcl}
      $H$ & $N=2$ hyper & & \\ \hline
      $\Psi_{L1}$, $\Psi_{R1}$ & $N=1$ & $\to$ & power running\\
      $\Psi_{L2}$, $\Psi_{R2}$ & $N=2$ hyper & $\to$ & logarithmic
      running\\
    \end{tabular}
    \caption{The suppression of anomalous dimensions due to the absence of
      Yukawa couplings among hypermultiplets.}
  \end{center}
\end{table}

\subsection{$N=2$ vector multiplet}

In this subsection, we discuss the possibility that the KK
contributions to the anomalous dimension $\gamma_H$
(Fig.~\ref{fig:higgs}) cancel out, i.\/e.\/\ the condition (ii) stated
in the beginning of this section. Recall that in models with $N=2$
SUSY, hypermultiplets receive no wave function renormalization.  This
is because Yukawa couplings are forbidden among hypermultiplets and in
addition, the radiative corrections from an $N=2$ vector multiplet are
cancelled between $N=1$ vector and chiral multiplets. We use this
non-renormalization property to suppress the Higgs anomalous
dimension.

Let us consider a two-generation model. We assume that ${\Psi_L}_1$ is
originated from an $N=2$ vector multiplet. That is, in higher
dimension, the standard gauge symmetry is enhanced to a larger gauge
group $G$ and broken by some symmetry (projection), leaving
${\Psi_L}_1$ as a part of broken gauge multiplet. In this case, we
should suppose that the Higgs field $H$ and ${\Psi_R}_1$, which are
combined with ${\Psi_L}_1$ into a Yukawa coupling, are originated from
an $N=2$ hypermultiplet. Note that above the KK threshold, since the
${\Psi_L}_1$ field belongs to the $N=2$ vector multiplet of the gauge
group $G$, $H$ and ${\Psi_R}_1$ are in a single representation of $G$. 
With this assumption, there is no radiative corrections in the Higgs
anomalous dimension. The anomalous dimension $\gamma_{R1}$ is also
suppressed while $\gamma_{L1}$ receives large contributions from KK
modes.

Now there are two options for the assignment of the second-generation
fields. One is to consider ${\Psi_{L,R}}_2$ have no KK modes (live
only in four-dimensional spacetime). Then no KK contribution to
$\gamma_H$ appears, and moreover the anomalous dimensions
${\gamma_{L,R}}_2$ are also zero due to the charge conservation law
discussed before. The other choice is that we assume the
${\Psi_{L,R}}_2$ fields belong to $N=2$ hypermultiplets and have no
Yukawa coupling. After all, in both cases, we find no enhanced
behaviors of the RG evolution for the second-generation Yukawa
coupling. We thus have a hierarchy between two Yukawa couplings.

The appearance of suppressed Yukawa coupling depends on whether it is
originated from $N=2$ vector multiplet. The mechanism can hence
generate a single mass gap between generations.
\begin{table}[htbp]
  \begin{center}
    \begin{tabular}{c|lcl}
      $H$ & $N=2$ hyper & & \\ \hline
      ${\Psi_L}_1$ & $N=2$ vector & & \\
      ${\Psi_R}_1$ & $N=2$ hyper & $\to$ & power running\\
      ${\Psi_L}_2$, ${\Psi_R}_2$ & $N=2$ hyper or $N=1$ & $\to$ &
      logarithmic running\\
    \end{tabular}
    \caption{The suppression of anomalous dimensions due to the vanishing
      anomalous dimensions of hypermultiplets.}
  \end{center}
\end{table}

\subsection{Constraint on gauge contribution}

So far, we have considered generic fixed-point values of Yukawa
couplings and their suppression mechanisms. If one applies these
mechanisms to the observed fermion mass hierarchy, one should impose
an additional phenomenological constraint on the beta-function
coefficients (the exponents of $\Lambda/\mu$). That is, the
fixed-point values must be of order one or less because the $SU(3)$
gauge and top Yukawa couplings are of order one and the other Yukawa
couplings are smaller than them. It is found from the fixed-point
solutions discussed in section 2 that this phenomenological
requirement results in the suppression of KK gauge contributions as
seen below.

Since all the KK modes contribute to the gauge anomalous dimension, we
naively have $\delta_g\geq\delta^g_i$. If this is the case, it follows
from the fixed-point solutions that the leading terms turn out to be
$\alpha_{y_i}\sim
(\mu_0/\Lambda)^{\delta^y_i-\delta_g}\alpha$. However, with the
requirement of gauge invariance, vector multiplets must spread into
the extra dimensions in which charged fields live. That may imply
$\delta_g\geq\delta^y_i$ and the available values of Yukawa couplings
become larger than order one: $\alpha_{y_i}\sim
(\mu_0/\Lambda)^{c_i}\alpha$ with $c_i<0$. If one identifies $\alpha$
with the $SU(3)$ gauge coupling, one can only have Yukawa couplings
larger than the top Yukawa coupling on the fixed point. We are thus
led to a phenomenological requirement that the exponent $\delta_g$
should be smaller than those in the matter anomalous dimensions. It is
interesting that this condition is favorable for Yukawa couplings to
realize a rapid convergence into the infrared fixed points.

One way to satisfy the above condition is to require the coefficient
of $(\Lambda/\mu)^{\delta_g}$ term vanishes. The exponent of the gauge
anomalous dimension is then reduced to a smaller one. Generally, the
gauge beta function is determined by the gauge and matter loop
diagrams (Fig.~\ref{fig:gauge}). In the matter loops, all gauge
non-singlet fields can propagate and contribute to the anomalous
dimensions. In this case, the beta function is expressed in terms of
$N=1$ basis,
\begin{eqnarray}
  \beta(\alpha) &\propto& -3C_2(G)\left(\frac{\Lambda}{\mu}\right)^{\delta^V}
  +\sum_{M} T(R_M)\left(\frac{\Lambda}{\mu}\right)^{\delta^M}
  +\sum_{M'} T(R_{M'})\left(\frac{\Lambda}{\mu}\right)^{\delta^{M'}}
  +\cdots
  \label{b},
\end{eqnarray}
where $C_2(G)$ is the quadratic Casimir and $T(R)$ denotes the Dynkin
index of the representation $R$. In the above, the first term is the
contribution of $N=1$ vector (and ghost) fields and the following part
comes from that of the KK matter fields
($\delta^M>\delta^{M'}>\cdots$).  The exponent $\delta_g$ in the beta
function is fixed by $\delta^V$, $\delta^M$, and so on. Suppose that
the first two terms in Eq.~(\ref{b}) cancel, we have a possibility of
$\delta_g<\delta^y_i$.  An interesting by-product of this cancellation
is that it is preferable to infrared fixed-point behavior. That is, in
the beta-function, the negative contribution from gauge fields is
cancelled and only the positive matter contributions remain. The
resultant infrared-free behavior of gauge coupling enhances the rate
at which Yukawa couplings approach to the fixed points~\cite{anf}.

We make a remark on the naturalness of this cancellation. If we regard
the models as low-energy effective descriptions of higher-dimensional
theories, the fields extended into higher dimension seem relevant to
larger supersymmetry, e.g.\ $N\geq 2$ SUSY\@. It is hence possible
that the leading power-law contributions from such fields cancel out
because of the larger symmetry. This line of thought could justify a
realization of finiteness condition, starting from a superstring
theory~\cite{HS}. Such a kind of condition generally imposes severe
restrictions in constucting realistic models.

We could further consider more generic situations where various gauge
fields live in various dimensions, that is, we have different gauge
group in each step of compactifications. In this case, the
beta-function coefficient is written as follows:
\begin{eqnarray}
  \beta(\alpha) &\propto& \left\{-3C_2(G_V)+\sum_{M}T(R_M)\right\}
  \left(\frac{\Lambda}{\mu}\right)^{\delta_V} \nonumber \\ &&+\left\{ -
  3C_2(G_{V-1})+\sum_{M'}T(R_{M'})\right\}
  \left(\frac{\Lambda}{\mu}\right)^{\delta_V-1} +\cdots.
\end{eqnarray}
Here, $G_x$ denotes the gauge group in the higher dimension ($G_V
\supset G_{V-1} \supset \cdots$). If the coefficients of power-law
terms vanish in each curly bracket, the resultant $\delta_g$ is quite
reduced and we could have hierarchical structures of Yukawa fixed
points between larger number of generations. This type of assumption
also has strong constraints on model building.

\section{Generation mixing}
\setcounter{equation}{0}

Until now, we have assumed that the Yukawa matrix takes a diagonal
form. In this section, we analyze further the fixed-point structure of
Yukawa matrices including the generation-mixing elements. We show
that, under Conditions I and II in section 2, a hierarchical structure
of Yukawa matrices is realized even if we include generation mixing
effect. As an example, we here take the scenario presented in section
3.1.

\subsection{Fixed-point structures}

First, we deal with a two-generation case for simplicity. Let us
suppose the following Yukawa couplings,
\begin{eqnarray}
  W &=& y_{ij} {\Psi_L}_i {\Psi_R}_j H,\qquad (i,j=1,2).
\end{eqnarray}
In addition, we assume that the `left-handed' fields $\Psi_L$ and
Higgs $H$ can propagate through the extra dimensions while, the
`right-handed' fields $\Psi_R$ reside in our four-dimensional
world. All the Yukawa couplings $y_{ij}$ are expected to take $O(1)$
values at the high-energy scale $\Lambda$. As explained in section
3.1, with this configuration the anomalous dimension of the Higgs
field receives only logarithmic corrections from the Yukawa terms due
to the charge conservations associated with the extra dimensions. On
the other hand, we obtain the anomalous dimensions of $\Psi_{L,R}$
neglecting sub-dominant corrections,
\begin{eqnarray}
  {\gamma_L}_{ij} &=& \frac{c_i}{16\pi^2}\, g^2
  \left(\frac{\Lambda}{\mu}\right)^{\delta^g_i}\delta_{ij}, \\
  {\gamma_R}_{ij} &=& -\sum_{k=1,2} \frac{a_k}{16\pi^2}\, (y^{\rm
  T})_{ik}\, y_{kj}\, \left(\frac{\Lambda}{\mu}\right)^{\delta_k}.
\end{eqnarray}
The anomalous dimension $\gamma_L$ receives the KK contributions from
gauge multiplets but not from the Yukawa terms, $y_{ij}$. As for
$\gamma_R$, the opposite situation occurs. Each contribution
corresponds to the diagram depicted in Fig.~\ref{fig:matter}.  In the
above expressions of $\gamma$'s, the coefficients $a_i$ and $c_i$ are
order one quantities including the group indices and the volume
factors. In the following, we will neglect these coefficients as
before.

To see the infrared fixed-point structure, let us write down the
beta-functions of the Yukawa couplings: 
\begin{eqnarray}
  \frac{d y_{ij}}{d t} &=& \frac{1}{16\pi^2} \beta_{ij} \;=\; 
  \sum_k \left(\gamma_{Lik}\,y_{kj}+y_{ik} \gamma_{Rjk} \right),
\end{eqnarray}
or more explicitly,
\begin{eqnarray}
  \beta_{11} &=& \left[y_{11} g^2 
    \left(\frac{\Lambda}{\mu}\right)^{\delta^g_1} 
    -y_{11}\left(y_{11}^2+y_{12}^2\right)
    \left(\frac{\Lambda}{\mu}\right)^{\delta_1} 
    -y_{21}\left(y_{11}y_{21}+y_{12}y_{22}\right)
    \left(\frac{\Lambda}{\mu}\right)^{\delta_2} \right], \nonumber\\
  \beta_{12} &=& \left[y_{12} g^2
    \left(\frac{\Lambda}{\mu}\right)^{\delta^g_1} 
    -y_{12}\left(y_{11}^2+y_{12}^2\right)
    \left(\frac{\Lambda}{\mu}\right)^{\delta_1} 
    -y_{22}\left(y_{11}y_{21}+y_{12}y_{22}\right)
    \left(\frac{\Lambda}{\mu}\right)^{\delta_2} \right], \nonumber\\
  \beta_{21} &=& \left[y_{21} g^2
    \left(\frac{\Lambda}{\mu}\right)^{\delta^g_2} 
    -y_{11}\left(y_{11}y_{21}+y_{12}y_{22}\right)
    \left(\frac{\Lambda}{\mu}\right)^{\delta_1} 
    -y_{21}\left(y_{21}^2+y_{22}^2\right)
    \left(\frac{\Lambda}{\mu}\right)^{\delta_2} \right], \nonumber\\
  \beta_{22} &=& \left[y_{22} g^2
    \left(\frac{\Lambda}{\mu}\right)^{\delta^g_2} 
    -y_{12}\left(y_{11}y_{21}+y_{12}y_{22}\right)
    \left(\frac{\Lambda}{\mu}\right)^{\delta_1} 
    -y_{22}\left(y_{21}^2+y_{22}^2\right)
    \left(\frac{\Lambda}{\mu}\right)^{\delta_2} \right].
\end{eqnarray}
The beta-functions are symmetric under an exchange of the
right-handed-type indices 1 and 2 because no particular assumption is
made on the right-handed fields $\Psi_R$. Though the beta-functions
seem to have complicated structures, as seen below we can find the
fixed-point solutions for several definite combinations of Yukawa
couplings. In what follows, we assume that the beta-function of the
gauge coupling do not have larger power of $(\Lambda/\mu)$ than
$\delta_g^{\chi_i}$. Note that this condition guarantees the strong
convergence of Yukawa couplings to the infrared fixed points~(see
section 2). In this case, neglecting lower-order terms of
$(\Lambda/\mu)$, the fixed-point solutions are derived from the Yukawa
beta-functions themselves without including the gauge beta functions.

Now consider the following combinations of Yukawa couplings:
\begin{eqnarray}
  X_1 \equiv y_{11}^2+y_{12}^2,\qquad X_2 \equiv y_{21}^2+y_{22}^2,
  \qquad Z \equiv y_{11}y_{21}+y_{12}y_{22}.
  \label{XZ}
\end{eqnarray}
The beta-functions for $X_{1,2}$ and $Z$ are given by
\begin{eqnarray}
  \frac{d X_1}{d t} &=& \frac{2}{16\pi^2} \left[
    \left(\frac{\Lambda}{\mu}\right)^{\delta_{g_1}}g^2X_1 
    -\left(\frac{\Lambda}{\mu}\right)^{\delta_1}X_1^2
    -\left(\frac{\Lambda}{\mu}\right)^{\delta_2} Z^2 \right],\\
  \frac{d X_2}{d t} &=& \frac{2}{16\pi^2} \left[
    \left(\frac{\Lambda}{\mu}\right)^{\delta_{g_2}}g^2X_2 
    -\left(\frac{\Lambda}{\mu}\right)^{\delta_2}X_2^2
    -\left(\frac{\Lambda}{\mu}\right)^{\delta_1} Z^2 \right],\\
  \frac{d Z}{d t} &=& \frac{1}{16\pi^2} Z \left[
    \left(\frac{\Lambda}{\mu}\right)^{\delta_{g_1}}g^2 
    +\left(\frac{\Lambda}{\mu}\right)^{\delta_{g_2}}g^2
    -2\left(\frac{\Lambda}{\mu}\right)^{\delta_1}X_1 
    -2\left(\frac{\Lambda}{\mu}\right)^{\delta_2}X_2 \right].
\end{eqnarray}
From these, we find the fixed-point solutions at the KK threshold scale $\mu_0$,
\begin{eqnarray}
  X_1^* \;=\;
  g^2\left(\frac{\Lambda}{\mu_0}\right)^{\delta_{g_1}-\delta_1},\quad
  X_2^* \;=\;
  g^2\left(\frac{\Lambda}{\mu_0}\right)^{\delta_{g_2}-\delta_2},\quad 
  Z^* \;=\; 0.
  \label{fpsol}
\end{eqnarray}
It is easily shown that these solutions are infrared stable against
small fluctuations about the solutions. We thus find a hierarchy
between the Yukawa couplings as the infrared fixed-point predictions
with one parameter $\theta$,
\begin{eqnarray}
  y_{ij} &\simeq& \pmatrix{
    x \cos\theta & x \sin\theta \cr
    -\sin\theta & \cos\theta \cr}
  \left(\frac{\mu_0}{\Lambda}\right)^{(\delta_2-\delta^g_2)/2} g,  \label{matrix}\\
x&\equiv&\left(\frac{\mu_0}{\Lambda}\right)^{(\delta^g_2-\delta_2-\delta_1^g+\delta_1)/2}.
\end{eqnarray}
The parameter $\theta$ and the relative sign in the matrix are to be
fixed by initial values of the Yukawa couplings. 

In this way, we find a particular form of the low-energy Yukawa matrix
as the fixed-point solutions. For this, we have only assumed that the
anomalous dimensions of the left-handed fields, $\gamma_L$, dominate
the beta-functions of Yukawa couplings. An interesting feature is that
on the fixed point the Yukawa matrix realizes a hierarchy of order $x$
even when the matrix elements do not have any hierarchies in initial
conditions. The hierarchy is induced by the RG evolution with
different values of $\delta_i$'s, i.\/e.\/\ the numbers of extra
dimensions which the corresponding fields ($\Psi_L$ in the present
case) can feel. After diagonalizing the matrix on the fixed point, one
obtains the eigenvalues with a ratio $y_1:y_2=x:1$, where the larger
eigenvalue is of order $(\mu_0/\Lambda)^{(\delta_2-\delta^g_2)/2}$.

Another interesting point is the mixing of generations. It is found
that the mixing angles of the left-handed fields ${\Psi_L}_i$ are
small in the infrared region. (In the particular form of matrix
(\ref{matrix}), the mixing angle of $\Psi_L$ ($=(Z/X_2)^2$) is zero.) 
\ This is a remarkable result: that gives a natural explanation why
the observed quark mixing angles are so small. The mixing angles are
expected to be of order one at high energy but they are reduced to
very small values under the RG evolution with large KK-mode
effects. More interestingly, the smallness of mixing angles is
realized on the infrared stable fixed point and does not depend on
details of models (symmetry properties, etc.). This novel explanation
of the small CKM angles has a geometrical origin and is different from
other approaches; the Higgs mixing, the higher-dimensional operators,
etc.

Exactly speaking, with a simple field configuration discussed above,
the mixing angle of the left-handed fields is almost zero on the
infrared fixed point. However, a more complicated structure, for
example, where the right-handed fields $\Psi_R$ also feel some number
of extra dimensions, can induce nonzero mixing angles between the
generations. We leave this issue to future investigations, and in this
paper, only do order-of-magnitude estimations of each matrix element.

We comment on the mixing angles of the right-handed fields
$\Psi_R$. With the matrix form (\ref{matrix}), the right-handed mixing
is given by $\tan\theta$. It cannot be settled on the infrared
fixed-point values but it is natural to expect that it is of order
one.  Note that the above analysis can be applied to the Yukawa
couplings of the fields with strong gauge interactions. When one
considers the grand unification scenarios, the lepton mixing angles
are also predicted by the gauge symmetry. In particular, in the
$SU(5)$ formalism, the right-handed down quarks and the left-handed
charged leptons are contained in a single multiplet $5^*$ and their
mixing angles are closely connected. Both mixing angles therefore
become large if one assumes that the behavior of down-quark Yukawa
matrix (i.\/e.\/\ $10\;5^*\,H_d$ couplings) is described by the above
mechanism. The large mixing angles give the boundary conditions at the
grand unification scale. This scenario could explain the large lepton
mixing recently observed in the Superkamiokande
experiment~\cite{superk}.

\subsection{A toy model}

We here adopt the mechanism for generating hierarchies among the
Yukawa couplings discussed in the previous section, and construct a
model (field configurations) which leads to the realistic up and
down-quark mass hierarchies. We only perform the order-of-magnitude
estimations.

The essential points of the mechanism are: (i) the left-handed fields
propagate through the extra dimensions in order to obtain mass
hierarchies besides the small generation mixing. (ii) the overlapping
of the left-handed fields with the Higgs field determines the sizes of
hierarchies. From (i), the configuration of the $SU(2)$ doublet quarks
$Q_i\,(i=1,2,3)$ sets the mass hierarchies between the generations. In
this case, one may naively wonder that the up and down quarks have
only the same order of hierarchies. However, notice that the
hierarchical factors are determined by the overlap of $Q_i$ and the
Higgs fields. That is, if the up-type Higgs $H_u$ and the down-type
one $H_d$ feel different numbers (different directions) of extra
dimensions, one can realize different hierarchical structure between
the up and down parts. Let us consider the field configuration shown
in Table~\ref{QH1}.
\begin{table}[htbp]
  \begin{center}
    \leavevmode
    \begin{tabular}{c|ccccccc}
      & ~5\quad & \quad 6\quad & \quad 7\quad & \quad 8\quad & \quad
      9\quad & \quad 10\quad \\ \hline
      $Q_1$ & ~$\bigcirc$ & \quad $\bigcirc$ & \quad $\bigcirc$ &
      \quad $\bigcirc$ & \quad $\bigcirc$ & \quad $\bigcirc$ \\
      $Q_2$ & ~$\bigcirc$ & \quad $\bigcirc$ & \quad $\bigcirc$ &
      & & \\
      $Q_3$ &  &  & & & & \\ 
      $H_u$ & ~$\bigcirc$ & \quad $\bigcirc$ & \quad $\bigcirc$ &
      \quad $\bigcirc$ & \quad $\bigcirc$ & \quad $\bigcirc$ \\
      $H_d$ & ~$\bigcirc$ & & \quad $\bigcirc$ & & \quad $\bigcirc$ & \\
    \end{tabular}
    \caption{An example of field configuration.}
    \label{QH1}
  \end{center}
\end{table}
We can easily extend the previous result of matrix form (\ref{matrix})
to the three generation case and find similar fixed-point structures
for several combinations of Yukawa couplings such as the determinant,
the mixing angles, the ratios, and so on. As a result, we obtain the
following hierarchical forms of the up and down Yukawa 
couplings at $\mu_0$;
\begin{eqnarray}
  y_u &\sim& \pmatrix{
    \epsilon^6 & \epsilon^6 & \epsilon^6 \cr
    \epsilon^3 & \epsilon^3 & \epsilon^3 \cr
    1 & 1 & 1 \cr}
  \epsilon^{-\delta_g'}g,\\[1mm]
  y_d &\sim& \pmatrix{
    \epsilon^3 & \epsilon^3 & \epsilon^3 \cr
    \epsilon^2 & \epsilon^2 & \epsilon^2 \cr
    1 & 1 & 1 \cr}
  \epsilon^{-\delta_g'}g,
\end{eqnarray}
where $\epsilon\equiv (\mu_0/\Lambda)^{1/2}$ and we have assumed all
the gauge contributions are same order (denoted by $\delta'_g$), for
simplicity. The Yukawa couplings are denoted in the basis of
$y_{ij}{\Psi_L}_i{\Psi_R}_j$ form. By diagonalizing the above
matrices, we obtain the mass hierarchies given by
\begin{eqnarray}
  m_u : m_c : m_t &\simeq& \epsilon^6 : \epsilon^3 : 1, \\
  m_d : m_s : m_b &\simeq& \epsilon^3 : \epsilon^2 : 1.
\end{eqnarray}
With an assumption $\epsilon\sim O(1/10)$, the obtained structures are
roughly consistent with the experimental data. Note that the
predictions are derived from the infrared fixed points and then
independent of the high-energy initial values. It is also interesting
that a larger hierarchy in the up-quark sector than in the down-quarks
is accomplished during the RG evolution.  This setup predicts a large
value of $\tan\beta$ and $\delta_g'=0$ for the top quark mass.

In this way, we can obtain various types of hierarchical forms of
Yukawa couplings on the infrared fixed points under the RG evolution
with the power-running effects. The essential point to have various
hierarchies is how the fields in the models spread in the extra
dimensions.

\section{Summary and discussion}
\setcounter{equation}{0}

In this paper, we have investigated a way to realize the generation
mass hierarchies in models with extra spatial dimensions. We have
exploited the enhanced (power-law) RG running effects originated from
KK excited modes. The power-law running effects significantly change
the behaviors of Yukawa couplings and induce hierarchically different
order of couplings as infrared fixed-point values. It is interesting
to determine precise low-energy values of couplings from the fixed
points since the predictions become independent of the initial
condition at high-energy scale. When applying this approach to
reproduce realistic spectrum, we have found several requirements to be
satisfied. The clarification of these conditions is one of the main
parts of this paper. First, the definite infrared fixed-point solution
should exist.  That gives several constraints on the beta-function
coefficients and hence restricts the possible forms of
higher-dimensional models. Second, the fixed-point values should be
flavor-dependent. It turns out to be a rather non-trivial problem to
obtain the flavor-dependent and suppressed Yukawa couplings. Naive
constructions of the models result in flavor-independent
(`democratic') type of mass matrices. This is mainly because the Higgs
anomalous dimensions enter in all the Yukawa beta-functions. We have
presented several mechanisms to avoid this problem and given a simple
example for each case.

We have also studied about the generation-mixing Yukawa
couplings. Under the RG evolution down to compactification scale, the
off-diagonal elements of the Yukawa matrix receive power-law
corrections from KK-mode contributions. We have analyzed a
two-generation case by utilizing one of the suppression mechanisms
presented in section 3. There only the left-handed fields as well as
the Higgs field are assumed to live in the extra dimensions while the
right-handed fields do not. We have shown that this assumption indeed
generates mass hierarchies on the stable fixed points. More
interestingly, the low-energy mixing angles of the left-handed fields
become very small even when their initial values are large at
high-energy region. This is due to the fact that the fields spread
into the extra dimensions and therefore can never be accomplished in
the usual four-dimensional scenarios. If we apply it to the quark
sector, we can explain why the observed CKM angles are very small.  On
the other hand, the right-handed mixing angles are of order one. When
one imagines a grand unified framework, the large lepton mixing, which
has recently been observed, is related to the quark sector and could
be explained in the same way.

In addition to the Yukawa hierarchy problem, supersymmetric models
generally involve another flavor problem coming from sfermion
mixing. They induce new sources of flavor violation such as
flavor-changing neutral currents and may give severe additional bounds
on the flavor structure of models.  However, it is recently shown in
Ref.~\cite{align} that when all Yukawa couplings are fixed by their
fixed-point values as supposed in this paper, the squark flavor
structure is aligned with that of the quarks and no additional flavor
problem arises.

Though we have not constructed explicit models, the obtained criteria
give some hints in constructing realistic models.
In particular, stringy construction is interesting.
For example, an orbifold model in general, includes several sectors, 
$N=4$ untwisted sectors, and $N=2$ and $N=1$ twisted sectors.
Such stringy model could provide with our set-up, and also give 
constraints, that is, e.g. in heterotic orbifold models 
the $N=4$ untwisted sector corresponds to $\delta =6$ extra 
dimensions and an $N=2$ sector has KK modes of $\delta =2$ extra 
dimensions, while the $N=1$ twisted sector has no KK tower, 
i.e.~$\delta =0$.
Furthermore, the beta function of the gauge coupling 
has no correction due to the $N=4$ sector.
This fact gives one type of realization $\delta_V = \delta_M$ and 
cancellation of these coefficients in Eq.~(3.1).
Similarly D-brane models could realize our set-up and 
also give constraints on configurations of extra dimensions $\delta$, 
which each field can feel.
We will study derivation of realistic Yukawa matrix from 
explicit combinations of $N=4, 2$ and $1$ sectors elsewhere.

\subsection*{Acknowledgments}

The authors would like to thank T.~Kugo for valuable discussions and
comments. M.~B.~is supported in part by the Grants-in-Aid for
Scientific Research No.~12047225(A2) and 12640295(C2) from the
Ministry of Education, Science, Sports and Culture, Japan.

\end{document}